\documentclass[
 reprint,
 amsmath,
 amssymb,
 aps,
 pre
]{revtex4-2}

\usepackage{graphicx}% Include figure files
\usepackage{dcolumn}% Align table columns on decimal point
\usepackage{bm}% bold math

\begin{document}

\title{Effect of expanding learning scope \\on the evolution of cooperation in scale-free networks}

\author{Masaaki Inaba}
\affiliation{Graduate School of Systems and Information Engineering, University of Tsukuba, Tsukuba, Ibaraki 305-8573, Japan}
\author{Eizo Akiyama}
\affiliation{Graduate School of Systems and Information Engineering, University of Tsukuba, Tsukuba, Ibaraki 305-8573, Japan}

\begin{abstract}
We study how expanding the scope for selecting game and learning (adaptation) partners affects the evolution of cooperation in public goods games on scale-free networks.
We show the following three results.
(i) Expanding the scope for selecting game partners suppresses cooperation.
(ii) Expanding the scope for selecting learning partners promotes cooperation when cooperation evolution is difficult.
(iii) When cooperation is more likely to evolve, slightly expanding the scope for selecting learning partners causes a significant drop in the cooperation rate, but expanding the scope further causes the cooperation rate to recover.
Although (i) is explained by the hub-centered mechanism, the well-known dynamic that promotes cooperation on scale-free networks, (ii) and (iii) are caused by a completely different mechanism that has heretofore been rarely mentioned.
\end{abstract}

\keywords{PACS number: 89.75.Hc, 87.23.Kg, 87.23.Ge, 02.50.Le}

\maketitle

\section{\label{sec:intro}Introduction}

There have been meaningful discussions \cite{Nowak2006,Rand2013,ZAGGL2014} of mechanisms to promote cooperation in social dilemma situations.
Some of the major mechanisms are kin selection, direct reciprocity, indirect reciprocity, network reciprocity, and multilevel selection.
Mechanisms other than network reciprocity often work under the assumption of well-mixed populations (random interactions) that are easy to model and analyze.
However, interactions among agents in the real world are rarely well-mixed, but rather resemble complex network structures \cite{Goh2002,Barabasi2016,Liu2021}.
In the study of network reciprocity, the mechanism of evolving cooperation has been studied using spatial structures such as circle graphs and lattices \cite{Nowak1992,Roca2009}, but these spatial structures cannot represent the network degree heterogeneity that is often observed in real-world human relationships.
Therefore, research on the evolution of cooperation in scale-free networks has augmented in recent years, based on the heterogeneous characteristics of the real world \cite{Santos2005,Ohtsuki2006,Rong2007,Santos2006,Santos2008,Apicella2012,Allen2017}.

The basic dynamics of the evolution of cooperation in scale-free networks have already been elucidated \cite{Santos2005,Ohtsuki2006,Santos2006,Santos2008}.
The heterogeneity of the network degree and the asymmetry of the cooperator and defector explain the dynamics.
Heterogeneity indicates a power-law distribution of the degree of each node in a scale-free network; in other words, a node with many degrees gets more degrees.
The asymmetry between cooperators and defectors means that when cooperators form a community, they enjoy high payoffs by cooperating with each other, whereas when defectors form a community, there are no benefits.
On a scale-free network with these two characteristics, cooperators can stably obtain higher payoffs than defectors by forming communities around nodes with high degree. Therefore, the cooperation rate is much higher in scale-free networks than in well-mixed populations or regular graphs, other conditions being equal, simply because of the spatial structure. That is the basic mechanism of network reciprocity on scale-free networks.

There are a number of studies \cite{Mahmoud2012,Mahmoud2012a,Shutters2012,Zhou2009,Podobnik2019,Chen2007,Pena2011,Peleteiro2014} that examine the effects of spatial structure by incorporating other cooperation mechanisms studied in well-mixed populations and regular networks into the study of network reciprocity on scale-free networks.
These studies are valuable because they test certain cooperation mechanisms in a realistic setting.
Our study furthers this type of work by focusing on the scope of games and learning (adaptation).

\begin{figure}[htbp]
  \includegraphics[width=8.5cm]{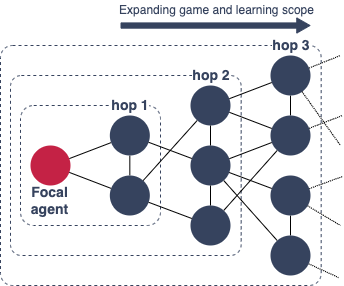}
  \caption{Scope of game and learning.\\Let the red node be a focal agent. We refer to a node whose shortest path distance to the focal agent is less than or equal to x as an agent in the hop x scope.}
  \label{fig1}
\end{figure}

Prior works on network reciprocity essentially assume that interactions, consisting of games and learning, only happen between neighboring agents in the network (hop 1 in Fig.\ref{fig1}).
However, in the real world, we do not always interact only with proximate agents.
Depending on the type of activity, we may interact not only with nearby agents, but also with distant agents (hop 2 and above in Fig.\ref{fig1}).
This type of interaction with distant peers is known as "weak ties \cite{Granovetter1973}" in empirical network analysis in sociology.
Although there have been several theoretical and experimental studies examining how the evolution of cooperation is affected by the scope of interactions \cite{Jun2007,Choi2008,Zhang2020,Okada2021}, there was insufficient focus on scale-free networks.
We investigate the effect of expanding the scope of interaction on the evolution of cooperation in public goods games on scale-free networks.

\section{\label{sec:method}Method}

We simulate public goods games (PGG) that consider the scope of interactions (games and learning) on a scale-free network with a multi-agent model based on evolutionary game theory.
PGG extends the prisoner's dilemma game to multiple players.
PGG is a natural representation of social dilemmas and has been adopted in many studies \cite{ElinorOstrom1990,Santos2008,Cao2010,Wang2012,Deng2016,Duh2020}.

The population structure is a static scale-free network generated by the Barabási–Albert (BA) model \cite{Barabasi1999}, with $N = 10^3$ nodes and the average degree $k = 4$.
Each node in the network represents an agent, and an agent’s strategy is either to cooperate (C) or to defect (D).
Initially, we randomly assign C or D with equal probability.
A generation consists of a game and a learning (adaptation), the games and learning for all agents are performed synchronously.

The game entails the following steps.
First, an agent is chosen as a focal agent.
Then, we randomly select $n - 1$ agents whose shortest path distance from the focal agent is within $hopG$ (game scope) as game partners.
Including the focal agent, the number of game partners is $n$.
The number of agents with strategy C among the n game partners is $n_C$.
Of the game partners, the agent with strategy C contributes cost $c = 1$, while the agent with strategy D pays no cost.
The total contribution is multiplied by $b$ and divided equally among all the game partners ($n$ players).
Thus, the payoff for each C agent is $\pi_C = \frac{n_C \times c \times b}{n} - c$, and the payoff for each D agent is $\pi_D = \frac{n_C \times c \times b}{n}$.
Each agent becomes the focal agent once and repeats this process to accumulate payoffs.
Note that the payoffs accumulated by each agent during the game are reset to zero at the beginning of each generation.

In the learning, as in the game, first, we assume that an agent is the focal agent.
Then, we randomly select $n - 1$ agents whose shortest path distance from the focal agent is within $hopL$ (learning scope) as the learning partners.
The agent with the highest payoff is chosen from among the learning partners, and if its payoff is greater than the focal agent's payoff, the focal agent copies the agent's strategy (C or D).
If the payoffs are the same or lower, the focal agent does not copy and does not change its strategy in the next generation.
In addition, the focal agent uses a random strategy in the next generation according to a mutation rate $\mu$.

\begin{figure}[htbp]
  \includegraphics[width=8.5cm]{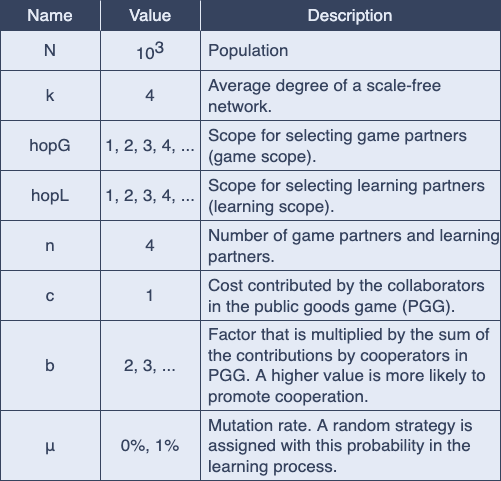}
  \caption{List of parameters.}
  \label{fig2}
\end{figure}

\section{Result}

\subsection{\label{sec:sim_result}Simulation result}

\begin{figure*}[htbp]
  \includegraphics[width=17cm]{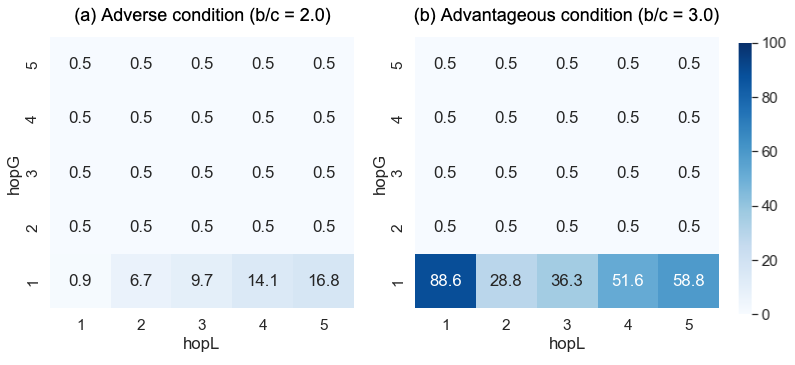}
  \caption{Population frequency of cooperators (\%).\\The x-axis ($hopL$) is the learning scope, the y-axis ($hopG$) is the game scope and each cell indicates the average population frequency of cooperators (\%) in $800 \sim 1000$ generations over 100 trials. The mutation rate ($\mu$) is 1\%.}
  \label{result1}
\end{figure*}

In the heat maps in Fig.\ref{result1}, the horizontal axis is the scope for selecting learning partners (learning scope, $hopL$) and the vertical axis is the scope for selecting game partners (game scope, $hopG$). The numbers in each cell represent the average population frequency of cooperators (cooperation rate) in the total population between 800 and 1000 generations over 100 trials.
We observe that the cooperation rate almost converges in the first 200 generations.
The heat map on the left corresponds to $b/c = 2.0$ and the right one corresponds to $b/c = 3.0$.
The setting on the right is more likely to evolve cooperation than that on the left.
The figures indicate the following three results.

(i) Expanding the game scope suppresses cooperation.
The cooperation rate converges to approximately $0.5\%$ when $hopG$ is greater than or equal to two, under both settings $b/c = 2.0$ and $3.0$.

(ii) Expanding the learning scope promotes cooperation when cooperation is difficult to evolve.
Under the adverse condition ($b/c = 2.0$), the cooperation rate is $0.9\%$ when $(hopG, hopL) = (1, 1)$, indicating that almost no cooperator survives. However, the cooperation rate is $16.8\%$ when $(hopG, hopL) = (1, 5)$, meaning that the cooperation rate improves as the learning scope ($hopL$) expands.
On a scale-free network, $hopL = 5$ is approximately equivalent to a well-mixed population, and the results do not change even if $hopL$ is increased further.

(iii) If cooperation is more likely to evolve, slightly expanding the learning scope causes a substantial drop in the cooperation rate, but expanding the scope further recovers the cooperation rate.
Under the advantageous condition ($b/c = 3.0$), the cooperation rate is $88.6\%$ when $(hopG, hopL) = (1, 1)$, indicating that many cooperators survive. However, slightly expanding the learning scope ($hopL = 2$) drops the cooperation rate significantly to $28.8\%$.Then, further expanding the learning scope ($hopL = 5$) recovers the cooperation rate to $58.8\%$.

We discuss the mechanisms that produce these results in the following sections.

\subsection{\label{sec:mecha1}Why does expanding game scope decrease the cooperation rate?}

As shown in Section \ref{sec:sim_result} (i), expanding game scope, regardless of $b / c$, suppresses the evolution of cooperation.
This phenomenon is particularly evident when we observe that at $b/c = 3.0$, where cooperation is relatively easy to evolve, $88.6\%$ of the agents are C when $(hopG, hopL) = (1, 1)$, while there are only $0.5\%$ C agents when $(hopG, hopL) = $(2, 1)$.
When $(hopG, hopL) = $(1, 1)$, the asymmetry between the cooperator and the defector is exploited to form and expand groups of C agents around the hub.
The asymmetry we are referring to here is that C enjoys high payoffs if it forms a group around the hub, while D gets no benefit from forming a group.
Now, if we increase $hopG$ from one to two, thereby expanding game scope, D can invade C's group and free-ride more easily.
As a result, C cannot form a robust group and is thus conquered by D.
Thus, the mechanism that promotes cooperation around the hub, as researched in previous studies \cite{Santos2005,Ohtsuki2006,Santos2006,Santos2008}, would be canceled when $hopG$ is 2 or higher, thereby suppressing cooperation.

\subsection{\label{sec:mecha2}Why does expanding learning scope increase the cooperation rate?}

\begin{figure*}[htbp]
  \includegraphics[width=17cm]{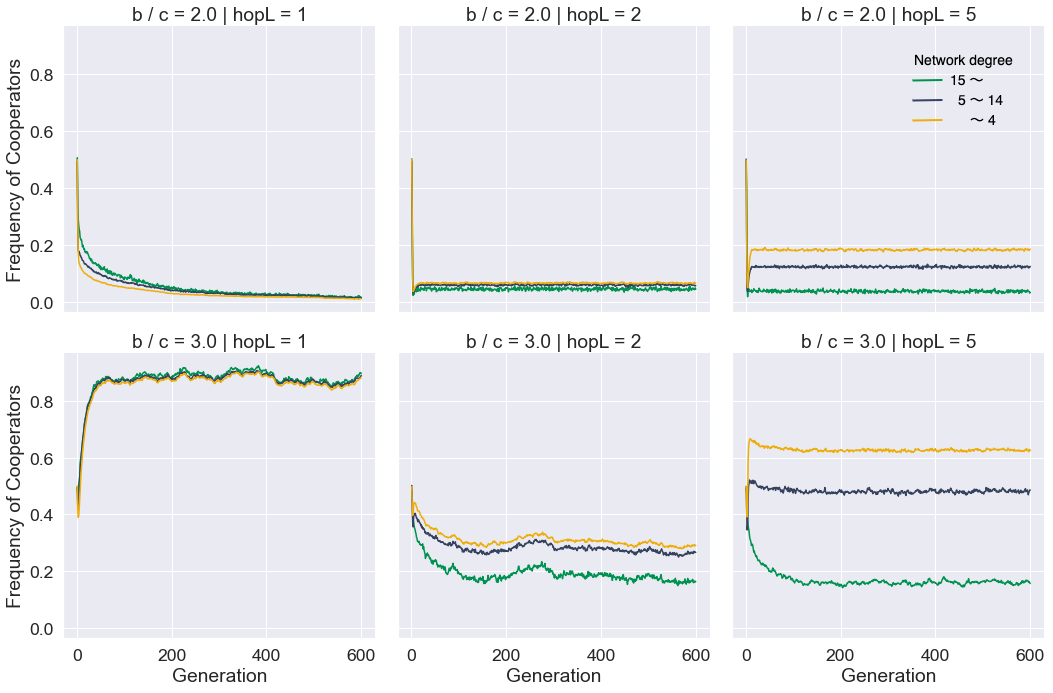}
  \caption{Population frequency of cooperators by network degree.\\The agents are classified into three network degree categories (high: $15\sim$, middle: $5\sim14$, low: $\sim4$). The transition of the frequency of cooperators in each category is shown with the generation on the horizontal axis. The three graphs in the upper row show $b/c = 2.0$, the three graphs in the lower row show $b/c = 3.0$, the graph in the left column shows $hopL = 1$, the graph in the middle column shows $hopL = 2$, and the graph in the right column shows $hopL = 5$. The number of trials is $100$, $hopG = 1$, and $\mu = 1\%$.}
  \label{result2}
\end{figure*}

Next, we examine why expanding the learning scope improves the cooperation rate, as shown in Section \ref{sec:sim_result} (ii).
We investigate this mechanism by categorizing each agent according to its network degree and looking at the cooperation rate, as shown in Fig.\ref{result2}.
Regardless of whether b/c is 2.0 or 3.0, when $hopL$ is one, the high-degree agents represented by the green line are slightly more cooperative than the low-degree agents represented by the orange line.
However, when $hopL$ is set to five, the cooperation rate of low-degree agents is higher than that of high-degree agents.
In other words, when the learning scope is narrow, high-degree agents lead the cooperative behavior, but when the learning scope expands, this is reversed, and low-degree agents appear to lead the cooperative behavior.
This suggests that expanding the learning scope promotes cooperation through a mechanism that is completely different from the hub-based mechanism that was well documented in previous studies \cite{Santos2005,Ohtsuki2006,Santos2006,Santos2008}.

\begin{figure*}[htbp]
  \includegraphics[width=18cm]{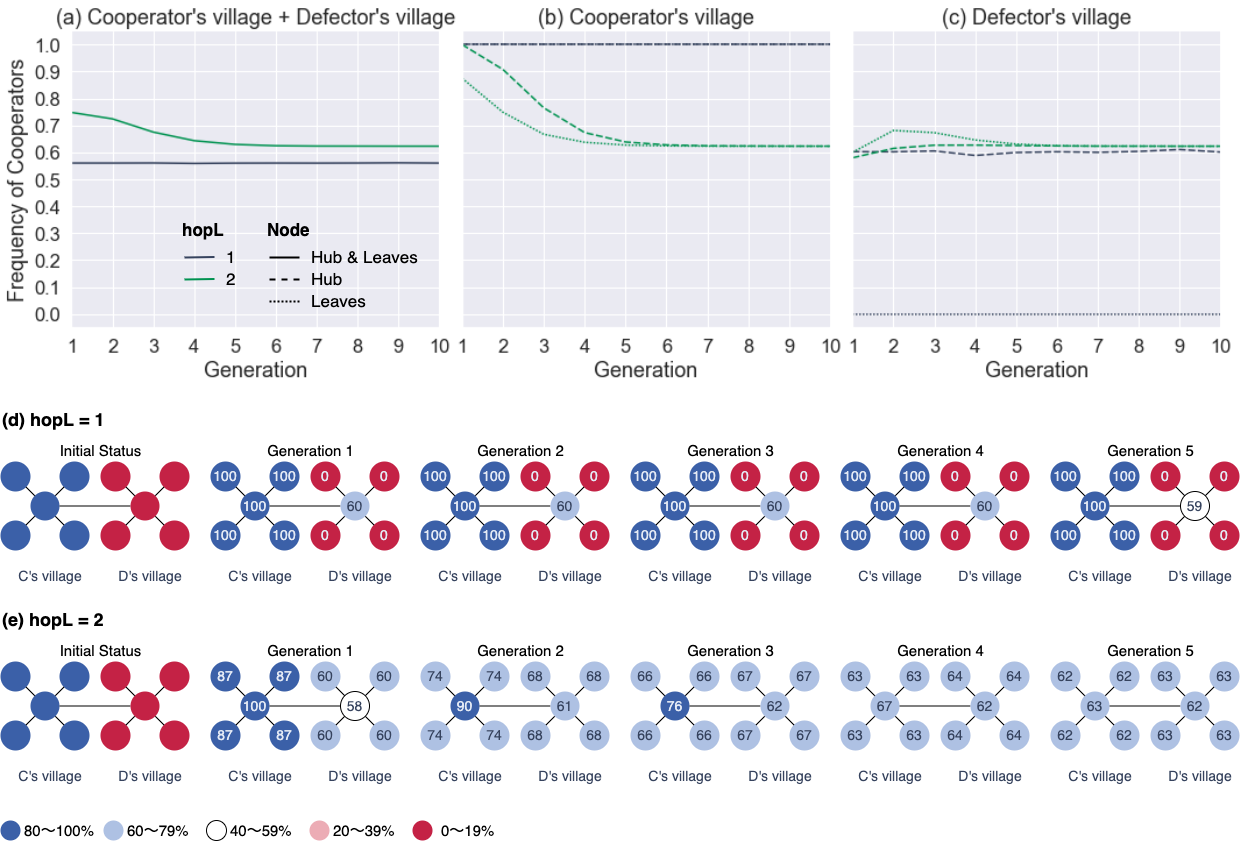}
  \caption{$hopL = 1$ v.s. $hopL = 2$ on mini-model ($b/c = 2.0$, $hopG = 1$)\\This figure shows the results of 10 generations of the simulation shown in "Section \ref{sec:method}: Method" using a mini-model with two star graphs connected ($10^3$ trials, $b = 2.0$, $n = 4$, $hopG = 1$, $\mu = 0\%$). The three line charts ((a), (b), (c)) show the transition of the frequency of cooperators in the whole population, the cooperator's village, and the defector's village, respectively, where the horizontal axis represents the generation. The color of the line indicates the learning scope, where gray is $hopL = 1$ and green is $hopL = 2$. The style of the lines represents the type of agent, with solid lines representing the sum of hubs and leaves, sparse dashed lines representing hubs, and dense dashed lines representing leaves. (d) and (e) show how each agent changes from its initial status to the fifth generation, comparing the case where $hopL = 1$ with the case where $hopL = 2$. The numbers in the circles indicate the probability that the agent becomes C in the next generation: the darker the blue, the higher the probability, and the darker the red, the lower the probability.}
  \label{mini1}
\end{figure*}

To examine this mechanism in more detail, we consider a mini-model as shown in Fig.\ref{mini1} (d) and (e).
This mini-model is an emphasized and deformed representation of the relationship between high-degree agents and low-degree agents in a scale-free network as a double star graph.
In the initial status, the agents in the left star are C and the agents in the right star are D.
We call the left side the cooperator's village and the right side the defector's village.
The three line charts in the Fig.\ref{mini1} (a), (b), and (c) show the results of a 10 generation, $10^4$ trial simulation based on this mini-model with the settings described in Section \ref{sec:method}: Method ($b/c = 2.0$, $hopG = 1$, $n = 4$, $\mu = 0\%$).

First, we examine the transition in the cooperation rate of the total population, which is the sum of the cooperator's village and the defector's village (Fig.\ref{mini1} (a)).
Compared with the gray line ($hopL = 1$), the green line ($hopL = 2$) is slightly higher, indicating that the phenomenon of increasing cooperation with expanding learning scope is also reproduced in this mini-model.

Next, we separate the total population (Fig.\ref{mini1} (a)) into the cooperator's village (Fig.\ref{mini1} (b)) and the defector's village (Fig.\ref{mini1} (c)).
In the cooperator's village, both gray lines (hopL = 1) maintain a 100\% cooperation rate from start to finish, whereas the green dense dashed line ($hopL = 2$, leaves) drops first, and then the green sparse dashed line (hopL = 2, hub) drops to approximately $60\%$.
In the defector's village (Fig.\ref{mini1} (c)), for hubs (sparse dashed lines), we see that there is no significant change in the cooperation rate of about $60\%$ even when $hopL$ increases from one (gray line) to two (green line).
In contrast, for the leaves (dense dashed line), when the $hopL$ is one (gray line), the cooperation rate is $0\%$, but when $hopL$ increases to two (green line), the cooperation rate is approximately $60\%$.
Because there are more leaves than hubs, the $60\%$ improvement for leaves has a greater overall impact.

Thus, expanding the learning scope lowers the cooperation rate in the cooperator's village and raises the cooperation rate in the defector's village.
The effect of increasing the cooperation rate of the leaves in the defector's village is greater than the decrease in the cooperator’s village; therefore, expanding the learning scope raises the cooperation rate as a whole.

This is further elucidated by examining the transition of the cooperation rate by agent (Fig.\ref{mini1} (d) and (e)).
Note that there is no significant difference in the hubs when comparing $hopL = 1$ with $hopL = 2$ at generation 1, but there is a significant change in the leaves.
This is because at $hopL = 1$, a leaf is only affected by the hub of the village to which it is directly connected, while at $hopL = 2$, it is also affected by the hub of the other village.
As a result, the overall cooperation rate settles at about $60\%$.
(See Supplemental Information for the analytical solution for generation 1)

\subsection{\label{sec:mecha3}Why does expanding learning scope decrease the cooperation rate?}

As Section \ref{sec:sim_result} (iii) shows, when $b/c = 3.0$ and we change $(hopG, hopL) = (1, 1)$ to $(1, 2)$, the cooperation rate drops significantly.
If, as \ref{sec:mecha2} shows, expanding the learning scope promotes cooperation, then increasing $hopL$ from one to two should increase the cooperation rate, but in fact it drops significantly.
We consider the mechanism that produces this phenomenon.
As in \ref{sec:mecha2}, a mini-model deforming the relationship between hubs and leaves in a scale-free network elucidates this mechanism.
The conditions are the same as in \ref{sec:mecha2} except that we increase b/c from 2.0 to 2.5.
Note that if we increase b/c to 3.0, the conditions become too favorable to cooperators and there is little decrease in the cooperation rate; if we increase it to 2.25, the cooperation rate does not converge in 10 generations.

\begin{figure*}[htbp]
  \includegraphics[width=18cm]{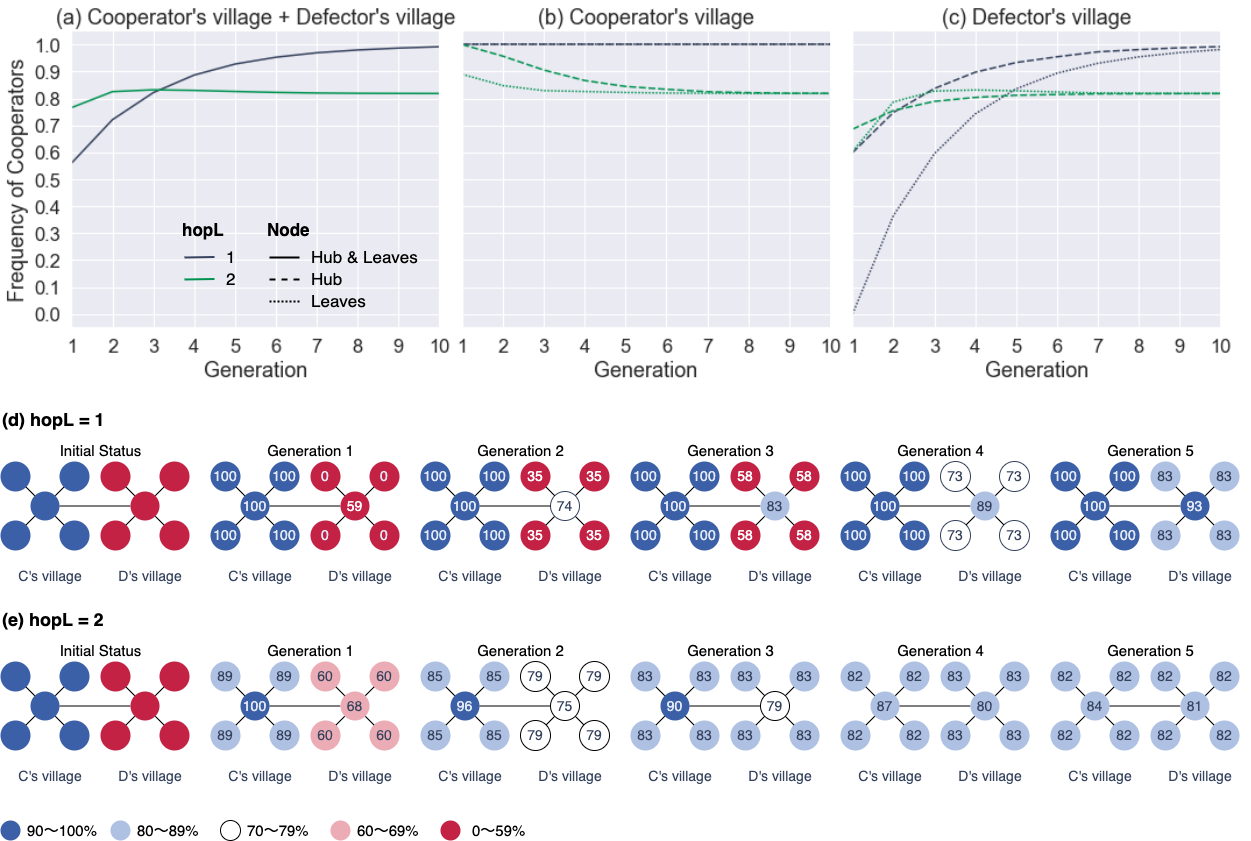}
  \caption{$hopL = 1$ v.s. $hopL = 2$ on mini-model ($b/c = 2.5$, $hopG = 1$)\\See Fig.\ref{mini1} for interpretation instructions; the difference is that we set $b/c$ to $2.5$ instead of $2.0$. The color scale is modified to emphasize the change in the cooperation rate.}
  \label{mini2}
\end{figure*}

Examining the cooperation rate of the entire population (Fig.\ref{mini2} (a)), the green line ($hopL = 2$) is almost $20\%$ lower than the gray line ($hopL = 1$) at generation 10.
The green lines for the cooperator's village (Fig.\ref{mini2} (b)) and the defector's village (Fig.\ref{mini2} (c)) also converge to a similarly lower cooperation rate of almost $20\%$ less than the gray lines.
Recalling \ref{sec:mecha2}, when $b/c = 2.0$, the cooperation rate decreased in the cooperator's village and increased in the defector's village, and the overall cooperation rate of the entire population increased.
However, when $b / c = 2.5$, the cooperation rate decreases in both villages, and thus the cooperation rate of the entire population decreases.

Next, we examine the transition diagram for each agent (Fig.\ref{mini2} (d) and (e)).
Comparing $hopL = 1$ and $2$ at generation 1, there is no significant difference in the probability of the hubs becoming C in the next generation, while the probabilities of the leaves becoming C have changed significantly.
This is because when $hopL = 1$, the leaves are only affected by the hub of the same village, whereas when $hopL = 2$, it is affected by the hub of another village.
As a result, the overall cooperation rate settles at over $80\%$.
(See Supplemental Information for an analytical solution for generation 1).

The mini-model is too small to reproduce the "trough" where the cooperation rate drops and then recovers, as seen in $(b/c, hopG, hopL) = (3, 1, 1-5)$ of Fig.\ref{result1}, even if the learning scope is further expanded.
However, two forces account for this "trough" phenomenon: the force that promotes cooperation by clustering cooperators around hubs, and the force that promotes cooperation by the expanding the learning scope.
That is, when $(b/c, hopG, hopL) = (3, 1, 1)$ in Fig.\ref{result1}, the hub force acts, and when $(b/c, hopG, hopL) = (3, 1, 5)$, the learning scope force acts.
These two forces cannot act at the same time because they originate from conflicting mechanisms, as Fig.4 suggests.
Hence, there is a "trough" between $(b/c, hopG, hopL) = (3, 1, 1)$ and $(3, 1, 5)$, where neither of the two forces is active when $(b / c, hopG, hopL) = (3, 1, 2)$ in Fig.\ref{result1}.

This brings us to the mechanism that produces the three results in Section \ref{sec:sim_result}.
We describe this mechanism as follows.
First, expanding both the game and the learning scope precludes the hubs from promoting cooperation.
Expanding the game scope will only suppress the hub's force, and nothing will happen even if the scope continues to expand.
In contrast, expanding the learning scope suppresses the hub's force at first, but then promotes cooperative behavior through another mechanism led by the leaves rather than the hubs.
When the hub’s force cannot act from the outset ($b/c = 2.0$), only the promotion of cooperation by expanding the learning scope becomes apparent. When the hub’s force can act ($b/c = 3.0$), the model demonstrates both the cancelation of the hub’s force and the promotion of cooperative behavior by expanding the learning scope.

\section{\label{sec:conclusion}Conclusion}

We studied the effect of expanding the scope of selecting game partners (game scope) and learning partners (learning scope) in public goods games on scale-free networks.
We demonstrate the following three results.
(i) Expanding game scope suppresses cooperation.
(ii) Under conditions limiting the evolution of cooperation, expanding learning scope promotes cooperation.
(iii) Under conditions where cooperation is more likely to evolve, slightly expanding learning scope causes a drop in the cooperation rate but expanding the learning scope further causes the cooperation rate to recover.
Previous studies on the evolution of cooperation in scale-free networks do not consider the scope of games and learning; our study is valuable in that it clarifies these results and the mechanisms by which they occur.
It is also important to point out that these results are caused by a completely different mechanism than the previously described promotion of cooperative behavior by hubs.

Future research should focus on further exploring some aspects of our study.
First, we consider that scale-free networks are realistic networks in the sense that they are more realistic than well-mixed populations and regular graphs; however, they are not always the best choice \cite{Broido2019}, and using other network structures is an interesting direction. In addition, the network structure used in our research amounts to a multiplex network, because changing the scope of the game and learning represents different network structures for the game and learning.
In this sense, we expect that this research relates to the accumulated research on multiplex networks \cite{Gomez-Gardenes2012,Wang2015,Pereda2016,Battiston2017,Allen2018}.
Second, although our study reveals the existence of mechanisms by which expanding the learning scope promotes and inhibits cooperation, our future work will involve further theoretical and analytical studies to identify and quantify the effects of specific parameters on the cooperation rate.
Further theoretical analysis of the relationship between scale-free networks and mini-models is also important.
Finally, this study shows the theoretical potential of the highlighted mechanisms. The dynamics of this mechanism in the real world must be verified and supported by research in other fields such as behavioral ecology and evolutionary anthropology.

\begin{acknowledgments}
We thank Irina Entin, M. Eng., from Edanz (https://jp.edanz.com/ac) for editing a draft of this manuscript.
\end{acknowledgments}

\bibliographystyle{unsrt}
\bibliography{Inaba2022_1}% Produces the bibliography via BibTeX.

\end{document}

\title{Supplemental Material for Effect of expanding learning scope on the evolution of cooperation in scale-free networks}

\maketitle

\onecolumngrid

\section*{Analytical solution of mini-model in I\hspace{-.1em}I\hspace{-.1em}I C}

Analytical solution for generation 1 of the mini-model in section I\hspace{-.1em}I\hspace{-.1em}I C is shown below. See "I\hspace{-.1em}I. Method" for the rules of the game and learning.

\begin{figure}[htbp]
  \includegraphics[width=17cm]{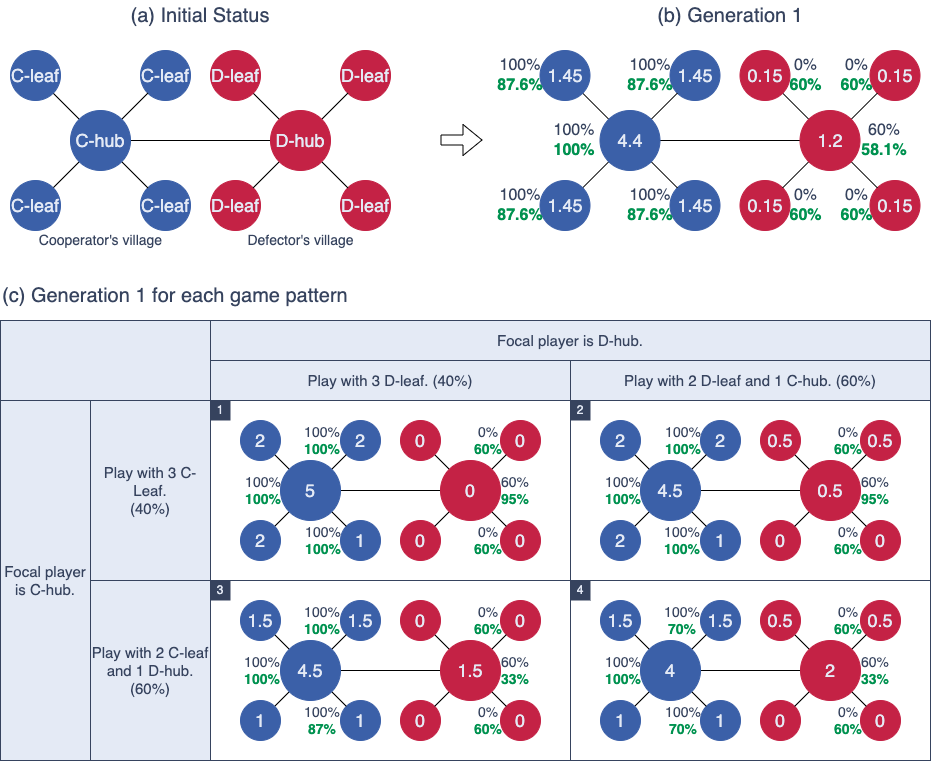}
  \caption{Analytical solution for generation 1 of the mini-model in section 3.3 (b/c = 2.0)\\b/c = 2.0, n = 4, hopG = 1, hopL = 1 and 2. The numbers in the circles represent the payoffs after generation 1, the black letters next to the circles represent the probability that the agent becomes C in generation 2 if hopL = 1, and the green letters next to the circles represent the probability that the agent becomes C in generation 2 if hopL = 2.}
  \label{fig:S1}
\end{figure}

There are two possible combinations of games in which the C-hub is the focal agent, and the probability of each occurring is shown in parentheses.

\begin{itemize}
	\item Played by 1 C-hub and 3 C-leaf.		(${}_4 C_3 / {}_5 C_3 = 40\%$)
	\item Played by 1 C-hub, 2 C-leaf and 1 D-hub.	(${}_4 C_2 / {}_5 C_3 = 60\%$)
\end{itemize}

Also, there are two possible combinations of games in which the D-hub is the focal agent, and the probability of each occurring is shown in parentheses.

\begin{itemize}
	\item Played by 1 D-hub and 3 D-leaf.		(${}_4 C_3 / {}_5 C_3 = 40\%$)
	\item Played by 1 D-hub, 2 D-leaf and 1 C-hub.	(${}_4 C_2 / {}_5 C_3 = 60\%$)
\end{itemize}

Therefore, there are four (2 × 2) patterns of game combinations in total, and cells 1 to 4 of fig. S1 (c).
For cell 1, calculate the payoffs at the end of generation 1 for each agent and the probabilities that each agent is C at the start of generation 2.
For the first step, calculate the payoff for each game.

\begin{itemize}
	\item Game 1		(when focal agent is C-hub):	(1 * 4) * 2 / 4 - 1 = 1
	\item Game $2\sim5$	(when focal agent is C-leaf):	(1 * 2) * 2 / 2 - 1 = 1
	\item Game 6	(when focal agent is D-hub):	(0 * 4) * 2 / 4 = 0
	\item Game $7\sim10$	(when focal agent is D-leaf):	(0 * 2) * 2 / 2 = 0
\end{itemize}

Therefore, the payoff for each agent is as follows.

\begin{itemize}
	\item C-hub:						5 (= 1 + 1 * 4)
	\item C-leaf (the 3 agents selected in game 1):	2 (= 1 + 1)
	\item C-leaf (the 1 agent not selected in Game 1):	1
	\item D-hub and D-leaf:				0
\end{itemize}

Next, calculate the probabilities that each agent is C at the start of generation 2 for hopL = 1.

\begin{itemize}
	\item C-hub and C-leaf:	100\%
	\item D-hub:			60\% ($= {}_4 C_2 / {}_5 C_3$)
	\item D-leaf:			0\%.
\end{itemize}

Finally, calculate the probabilities in the case of hopL = 2.

\begin{itemize}
	\item C-hub and C-leaf:	100\%
	\item D-hub:			95.24\% ($= 1 - {}_4 C_3 / {}_9 C_3$)
	\item D-leaf:			60\% ($= {}_4 C_2 / {}_5 C_3$)
\end{itemize}

Calculating as same for cells 2 to 4, we get

\begin{itemize}
	\item Cell 2
	\begin{itemize}
		\item Payoff for each game
		\begin{itemize}
			\item Game 1 (when focal agent is C-hub): (1 * 4) * 2 / 4 - 1 = 1
			\item Game $2\sim5$ (when focal agent is C-leaf): (1 * 2) * 2 / 2 - 1 = 1
			\item Game 6 (when focal agent is D-hub): (0 * 3 + 1 * 1) * 2 / 4 = 0.5\\(C-hub: -0.5, D-hub and D-leaf: 0.5)
			\item Game $7\sim10$ (when focal agent is D-leaf): (0 * 2) * 2 / 2 = 0
		\end{itemize}
		\item Payoff for each agent
		\begin{itemize}
			\item C-hub: 1 + 1 * 4 - 0.5 = 4.5
			\item C-leaf (the 3 agents selected in game 1): 1 + 1 = 2
			\item C-leaf (the 1 agent not selected in game 1): = 1
			\item D-hub: 0.5
			\item D-leaf (the 2 agents selected in game 6): 0.5
			\item D-leaf (the 2 agents not selected in game 6): 0
		\end{itemize}
		\item Probabilities that each agent is C in generation 2 when hopL = 1 are the same as in cell 1.
		\item Probabilities that each agent is C at the start of generation 2 when hopL = 2.
		\begin{itemize}
			\item C-hub $\&$ leaf: $100\%$
			\item D-hub: $1 - {}_4 C_3 / {}_9 C_3 = 95.24\%$
			\item D-leaf: ${}_4 C_2 / {}_5 C_3 = 60\%$
		\end{itemize}
	\end{itemize}
	\item Cell 3
	\begin{itemize}
		\item Payoff for each game
		\begin{itemize}
			\item Game 1 (when focal agent is C-hub): (1 * 3) * 2 / 4 = 1.5\\(C-hub and C-leaf: 0.5, D-hub: 1.5)
			\item Game $2\sim5$ (when focal agent is C-leaf): (1 * 2) * 2 / 2 - 1 = 1
			\item Game 6 (when focal agent is D-hub): (0 * 4) * 2 / 4 = 0
			\item Game $7\sim10$ (when focal agent is D-leaf): (0 * 2) * 2 / 2 = 0
		\end{itemize}
		\item Payoff for each agent
		\begin{itemize}
			\item C-hub: 0.5 + 1 * 4 = 4.5
			\item C-leaf (the 2 agents selected in game 1): 1 + 0.5 = 1.5
			\item C-leaf (the 2 agents not selected in game 1): 1
			\item D-hub: 1.5
			\item D-leaf: 0
		\end{itemize}
		\item Probabilities that each agent is C in generation 2 when hopL = 1 are the same as in cell 1.
		\item Probabilities that each agent is C at the start of generation 2 when hopL = 2.
		\begin{itemize}
			\item C-hub: $100\%$
			\item C-leaf (the 2 agents selected in game 1): $100\%$
			\item C-leaf (the 2 agents not selected in game 1): $1 - {}_4 C_3 / {}_5 C_3 / 3 = 86.67\%$
			\item D-hub: $1 - {}_8 C_3 / {}_9 C_3 = 33.33\%$
			\item D-leaf: ${}_4 C_2 / {}_5 C_3 = 60\%$
		\end{itemize}
	\end{itemize}
	\item Cell 4
	\begin{itemize}
		\item Payoff for each game
		\begin{itemize}
			\item Game 1 (when focal agent is C-hub): (1 * 3) * 2 / 4 = 1.5\\(C-hub and C-leaf: 0.5, D-hub: 1.5)
			\item Game $2\sim5$ (when focal agent is C-leaf): (1 * 2) * 2 / 2 - 1 = 1
			\item Game 6 (when focal agent is D-hub): (0 * 3 + 1 * 1) * 2 / 4 = 0.5\\(C-hub: -0.5, D-hub and D-leaf: 0.5)
			\item Game $7\sim10$ (when focal agent is D-leaf): (0 * 2) * 2 / 2 = 0
		\end{itemize}
		\item Payoff for each agent
		\begin{itemize}
			\item C-hub: 0.5 + 1 * 4 - 0.5 = 4
			\item C-leaf (the 2 agents selected in game 1): 0.5 + 1 = 1.5
			\item C-leaf (the 2 agents not selected in game 1): 1
			\item D-hub: 1.5 + 0.5 = 2
			\item D-leaf (the 2 agents selected in game 6): 0.5
			\item D-leaf (the 2 agents not selected in game 6): 0
		\end{itemize}
		\item Probabilities that each agent is C in generation 2 when hopL = 1 are the same as in cell 1.
		\item Probabilities that each agent is C at the start of generation 2 when hopL = 2.
		\begin{itemize}
			\item C-hub: $100\%$
			\item C-leaf: $1 - {}_4 C_2 / {}_5 C_3 / 2 = 70\%$
			\item D-hub: $1 - {}_8 C_3 / {}_9 C_3 = 33.33\%$
			\item D-leaf: ${}_4 C_2 / {}_5 C_3 = 60\%$
		\end{itemize}
	\end{itemize}
\end{itemize}

Fig.\ref{fig:S1} (c) shows these results. Fig.\ref{fig:S1} (b) shows the weighted average of the numbers in each cell according to the probability of occurrence. We can confirm that these results are consistent with generation 1 in Section I\hspace{-.1em}I\hspace{-.1em}I C Fig.5.

\section*{Analytical solution of mini-model in I\hspace{-.1em}I\hspace{-.1em}I D}

Analytical solution for generation 1 of the mini-model in section I\hspace{-.1em}I\hspace{-.1em}I D is shown below. The difference in the setting from the analytical solution of section I\hspace{-.1em}I\hspace{-.1em}I C is that b / c is changed from 2.0 to 3.0.

\begin{figure}[htbp]
  \includegraphics[width=17cm]{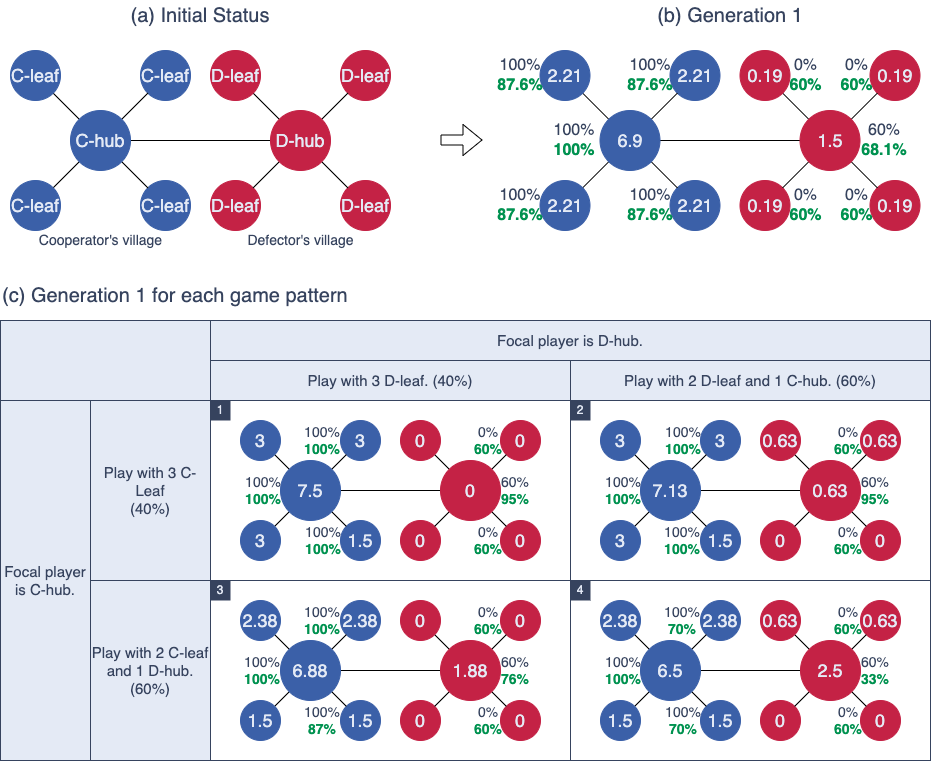}
  \caption{Analytical solution for generation 1 of the mini-model in section 3.4 (b/c = 2.5).\\b/c = 2.5, n = 4, hopG = 1, hopL = 1 and 2. The numbers in the circles represent the payoffs after generation 1, the black letters next to the circles represent the probability that the agent becomes C in generation 2 if hopL = 1, and the green letters next to the circles represent the probability that the agent becomes C in generation 2 if hopL = 2.}
  \label{fig:S2}
\end{figure}

The calculation process is described in the previous section. We can confirm that these results are consistent with generation 1 in section I\hspace{-.1em}I\hspace{-.1em}I D Fig.6.